\documentclass[pra, twocolumn,preprintnumbers,amsmath,amssymb,aps,pra]{revtex4}

\usepackage{mathrsfs}
\usepackage{graphicx}
\usepackage[caption=false]{subfig}
\usepackage{dcolumn}
\usepackage{amsmath}
\usepackage{epsfig}
\usepackage{color}
\usepackage{epstopdf}
\usepackage{framed}
\usepackage{float}
\usepackage{braket}
\begin{document}

\date{\today}

\title{Repeater-enhanced distributed quantum sensing based on continuous-variable multipartite entanglement}
\author{Yi Xia$^1$}
\email{yixia@email.arizona.edu}
\author{Quntao Zhuang$^3$}
\author{William Clark$^4$}
\author{Zheshen Zhang$^{2,1}$}
\address{1. College of Optical Sciences, University of Arizona, Tucson, Arizona 85721, USA\\
2. Department of Materials Science and Engineering, University of Arizona, Tucson, Arizona 85721, USA\\
3. Department of Physics, University of California, Berkeley, California 94720, USA \\
4. General Dynamics Mission Systems, 8220 East Roosevelt Street, Scottsdale, AZ 85257}%

\begin{abstract}
Entanglement is a unique resource for quantum-enhanced applications. When employed in sensing, shared entanglement between distributed quantum sensors enables a substantial gain in the measurement sensitivity in estimating global parameters of the quantum sensor network. Loss incurred in the distribution of entanglement, however, quickly dissipates the measurement-sensitivity advantage enjoyed by the entangled quantum sensors over sensors supplied with local quantum resources. Here, we present a viable approach to overcome the entanglement-distribution loss and show that the measurement sensitivity enabled by entangled quantum sensors beat that afforded by the optimum local resource. Our approach relies on noiseless linear amplifiers (NLAs) to serve as quantum repeaters. We show that unlike the outstanding challenge of building quantum repeaters to suppress the repeaterless bound for quantum key distribution, NLA-based quantum repeaters for distributed quantum sensing are realizable by available technology. As such, distributed quantum sensing would become the first application instance that benefits from quantum repeaters.

\end{abstract}


\maketitle

\section{\label{sec:level1}Introduction}
Quantum information science (QIS) gives rise to quantum cryptography \cite{gisin2002quantum}, quantum metrology \cite{degen2017quantum,giovannetti2006quantum}, and quantum computing \cite{bennett2000quantum,debnath2016demonstration,barends2016digitized}. These new paradigms offer functionalities beyond the reach of classical physics. For example, Shor's algorithm implemented on a quantum computer is able to exponentially reduce the computational processing of the best known classical algorithm for the prime factorization problem \cite{shor1999polynomial}. Quantum computing also allows for efficient simulation of quantum systems, opening a revolutionary route for the discovery of new materials \cite{georgescu2014quantum}. Based on the quantum no-cloning theorem and the Heisenberg uncertainty principle, quantum key distribution (QKD) enables unconditionally secure communication between distant parties. In a recent milestone experiment, QKD between two continents mediated by a low-earth-orbit satellite was achieved \cite{liao2018satellite}. QKD protocols capable of gigabit-per-second secret-key rates (SKRs) at metropolitan-area distances were also proposed \cite{zhuang2016floodlight} and demonstrated \cite{zhang2018experimental}. However, the fundamental rate-loss trade-off \cite{takeoka2014fundamental, pirandola2017fundamental} hinders QKD from obtaining gigabit-per-second SKRs at a global scale. Using quantum repeaters can mitigate the rate-loss trade-off, but building a quantum repeater that beats the fundamental QKD SKR bound on a lossy channel, known as the Pirandola-Laurenza-Ottaviani-Banchi (PLOB) bound, has been a formidable task. For example, Ref.~\cite{pant2017rate} shows that an all-photonic quantum repeater would require millions of ideal single-photon sources, something well beyond the reach of state-of-the-art technology.

Apart from quantum computing and quantum communication, the principles of QIS also promise measurement sensitivity beyond the standard quantum limit (SQL) \cite{giovannetti2006quantum}. Such an appeal has spurred considerable research effort on building quantum sensors that harness the quantum mechanical properties of light to enhance the measurement sensitivity while minimizing the illuminating power. Such a feature is critical for probing delicate objects such as biological samples that are susceptible to photobleaching or damage. A remarkable example is the application of the nonclassical single-mode squeezed vacuum (SV) light in the Laser Interferometer Gravitational-Wave Observatory (LIGO) to beat the SQL, i.e., the shot-noise level \cite{abadie2011a,aasi2013enhanced}. Single-mode SV light, however, only constitutes a subset of the class of nonclassical states of light. In this regard, multimode nonclassical states, in particular entanglement, are valuable resource for a variety of quantum-enhanced sensing applications. Specifically, entanglement was employed in quantum-enhanced microscopy \cite{ono2013entanglement}, the estimation of an optical phase \cite{xiang2011entanglement}, positioning systems and clock synchronization \cite{giovannetti2001quantum}, and magnetic field measurements \cite{sewell2012magnetic}. More recently, broadband entangled photons were utilized to increase the signal-to-noise ratio in interrogating the presence of a target embedded in a highly lossy and noisy environment that completely destroys the initial entanglement \cite{zhang2015entanglement}. Such a result indicates that quantum-enhanced sensing can be applied in practical scenarios. 

To date, most entanglement-enhanced sensing experiments have been dedicated to leveraging bipartite entanglement to increase the measurement sensitivity at a {\it single} sensor node. To fully unleash the power of entanglement, recent theoretical studies explored the use of entangled states in a distributed quantum sensing (DQS) setting \cite{proctor2018multiparameter, ge2018distributed, zhuang2018distributed}. Refs.~\cite{proctor2018multiparameter} and \cite{ge2018distributed} show that discrete-variable multipartie entangled (DVMP) states can substantially enhance the sensitivity of measuring the weighted sum of an unknown parameter. DVMP entangled states, however, are highly susceptible to loss, casting doubts on their merits in practical situations. In contrast, the quality of continuous-variable multipartite (CVMP) entanglement degrades graciously in the presence of loss. Indeed, Ref.~\cite{zhuang2018distributed} proved that the measurement-sensitivity advantage enabled by CVMP entanglement over nonclassical product states survives inefficient detectors, indicating that CVMP entanglement could empower robust quantum sensor networks for a wide range of {\it local} applications such as temperature  \cite{mehboudi2015thermometry} and angular momentum \cite{eckert2008quantum} measurements in cold-atom systems, probing pressure and stress induced refractive-index shifts, and noninvasive biomedical imaging \cite{eldredge2018optimal}. Nevertheless, for {\it nonlocal} applications such as long-baseline telescopes \cite{gottesman2012longer} and global-scale clock synchronization \cite{komar2014quantum}, loss arising from the distribution of CVMP entanglement to spatially separated sensor nodes quickly diminishes the measurement-sensitivity advantage over product-state DQS, because product states can be prepared locally without suffering the entanglement-distribution loss.

In this paper, we introduce a simple form of quantum repeater, the noiseless linear amplifier (NLA), into the nonlocal DQS regime to overcome the entanglement-distribution loss. We show that spatially separated quantum sensors sharing CVMP entanglement, in conjunction with NLAs after entanglement distribution, are able to beat the measurement sensitivity allowed by DQS enhanced by the {\it optimum} product states when both scenarios employ the same amount of probe power, but no such measurement-sensitivity advantage can be attained without NLAs. As such, NLAs used in CVMP-entanglement-based DQS effectively serve as quantum repeaters to mitigate entanglement-distribution loss. The proposed scheme would thus lead to the demonstration of the first quantum-enhanced sensing application that benefits from quantum repeaters.

The paper is organized as follows. In Sec.~\ref{sec:BuildingBlocks}, we overview the building blocks for the proposed DQS scheme. We first introduce CVMP-entanglement-based DQS and highlight its measurement-sensitivity advantage over DQS using the optimum product states. We then briefly describe the principle of the NLA. The main result of NLA-enhanced measurement-sensitivity for CVMP-entanglement-based DQS will be presented in Sec.~\ref{sec:MainResult}. The intuitions behind the main result will be discussed in Sec.~\ref{sec:Discussions}. Conclusion is presented in Sec.~\ref{sec:conclusion}.

\section{\label{sec:BuildingBlocks} Technical Background}

\subsection{\label{subsec:CVMP} Distributed Quantum Sensing based on CVMP Entanglement}
A rule of thumb for sensing is that multiple measurements on the same parameter improve the measurement sensitivity. By averaging the measurement outcomes, the root mean square (rms) estimation error, defined as the square root of the measurement sensitivity, scales as $1/\sqrt{M}$ where $M$ is the number of measurements. In a quantum sensor network, the rms error in estimating a global parameter by $M$ quantum sensors also scales as $1/\sqrt{M}$ if only product probe states are utilized. The $1/\sqrt{M}$ rms estimation error scaling is known as the standard quantum limit (SQL). Shared entanglement by different quantum sensors can be harnessed to beat the SQL. Recent theory works \cite{proctor2018multiparameter,ge2018distributed} show that discrete-variable multipartite (DVMP) entangled states give rise to a $1/\sqrt{M}$ improvement in the rms error in estimating a global parameter, e.g., the weighted sum of the magnetic field strength at different sensor nodes. DVMP entangled states, such as the GHZ-like states employed in Ref.~\cite{proctor2018multiparameter} and the two-Fock states discussed in Ref.~\cite{ge2018distributed}, have only been generated via post-selection, leading to a success probability that is exponentially small in $M$. Worse, DVMP entangled states are highly vulnerable to loss; the elimination of a few photons would completely destroy the entanglement and diminish the advantage over using product states. As such, DVMP entanglement-based DQS is formidable in practical situations.

\begin{figure}
    \centering
    \includegraphics[width=0.5\textwidth]{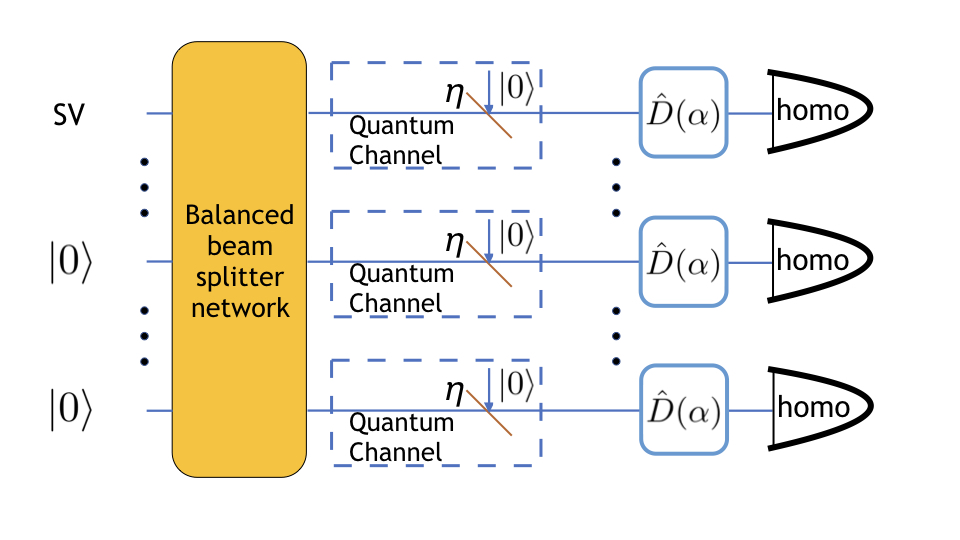}
    \caption{\label{fig:CVMP_DQS} Schematic of DQS based on CVMP entanglement. A squeezed vacuum (SV) state is mixed with $M-1$ vacuum modes on a balanced beam-splitter network to create CVMP entanglement as the probe state employed in DQS. Each mode of the entangled state is sent to a remote sensor node via a $\eta$-transmissvitity lossy channel. The physical parameter under estimation is a field displacement imparted by the operator $\hat{D}(\alpha)$. Homodyne measurements are performed at each sensor node, and their outcomes are averaged to obtain an estimate for $\alpha$. }
    \label{fig:distribute}
\end{figure}

CVMP entangled states, by contrast, can be generated deterministically via Gaussian operations \cite{weedbrook2012gaussian}. Moreover, the quality of CVMP entangled states degrades graciously in the presence of loss. Such nice properties of CVMP entangled states open a viable route toward DQS with practical constraints. Ref.~\cite{zhuang2018distributed} analyzed CVMP-entanglement-based DQS for the measurement of field displacement at the sensor nodes, as illustrated in Fig.~\ref{fig:CVMP_DQS}. A SV state with mean photon number $N_S$ mixes with $M-1$ vacuum modes on a lossless $M \times M$ beam splitter (BS) network. The output of the BS network is a CVMP entangled state, whose each mode serves as a sensor node to probe a uniform field displacement represented by the operator $\hat{D}(\alpha)$. The displaced modes are subsequently measured by homodyne detectors each with sub-unity quantum efficiency modeled by a $\eta$-transmissivity channel. The measurement sensitivity of CVMP-entanglement-based DQS is compared with that of the product-state DQS scheme in which each sensor node is supplied with a single-mode SV state with mean photon number $N_S/M$. The overall probe power for the CVMP-entanglement-based and the product-SV-state schemes are identical. Ref.~\cite{zhuang2018distributed} shows that with ideal homodyne detectors, i.e., $\eta = 1$, and per-mode mean photon number $n_S = N_S/M \gg 1$, the rms estimation error for the CVMP-entanglement-based DQS scheme scales as $\delta \alpha_{\eta = 1}^E \simeq 1/(4 M \sqrt{n_S})$, whereas the rms estimation error for product-SV-state quantum sensing scales as $\delta \alpha_{\eta = 1}^P \simeq 1/(4 \sqrt{M n_S})$. Evidently, the rms estimation error for CVMP-entanglement-based DQS beats the standard quantum limit. 

Ref.~\cite{zhuang2018distributed} further proved the optimality of the CVMP entangled state for DQS when the detection is restricted to homodyne measurements and demonstrated its measurement-sensitivity advantage over product-SV-state DQS even in the presence of detector inefficiency. Note that we will present in Appendix \ref{appendix:opt_proof} a general proof for the optimality of product-SV-state DQS without assuming any specific measurement scheme. Ref.~\cite{zhuang2018distributed} compared the measurement sensitivities of CVMP-entanglement-based DQS and product-CV-state DQS when the two scenarios are subject to equal loss. However, since SV states can be generated locally at each sensor node, they do not need to be distributed over a lossy channel. Thus, the measurement-sensitivity advantage enjoyed by CVMP-entanglement-based DQS over product-SV-state DQS will diminish quickly in the presence of entanglement-distribution loss. We will show in Sec.~\ref{sec:MainResult} that NLAs are able to overcome entanglement-distribution loss and regain the measurement-sensitivity advantage enabled by CVMP entanglement. Let us first brief review the principle of the NLA.

\subsection{\label{subsec:NLA} Noiseless Linear Amplifier}
\begin{figure}[htp]
  \includegraphics[width=0.5\textwidth]{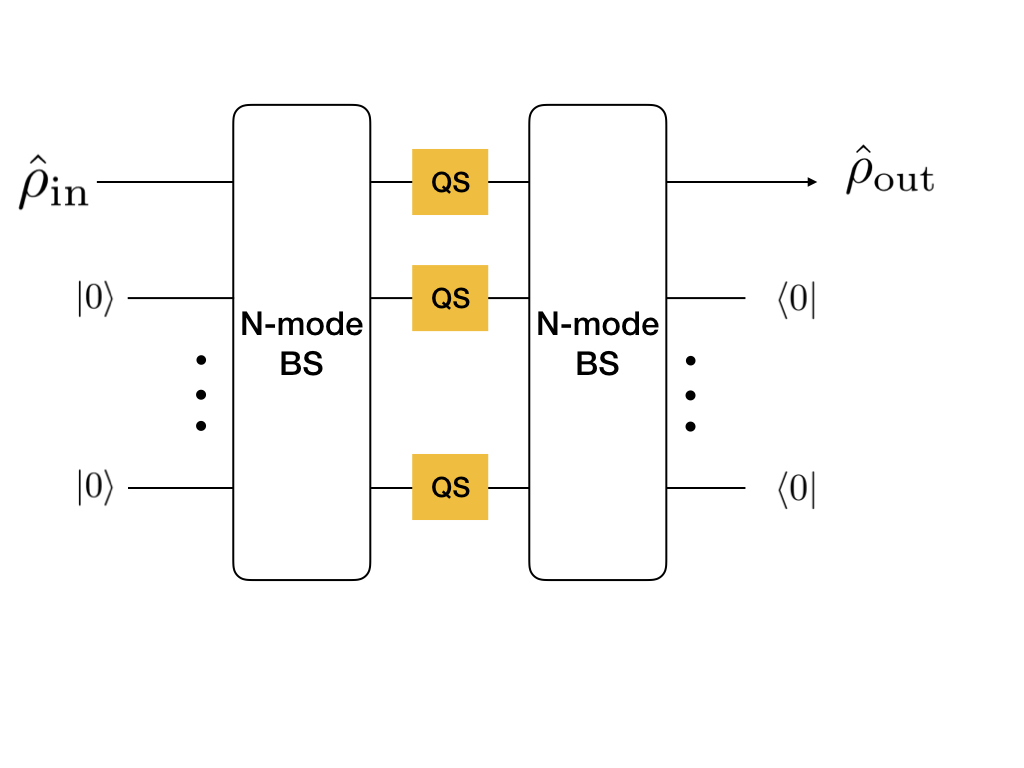}
  \includegraphics[width=0.5\textwidth]{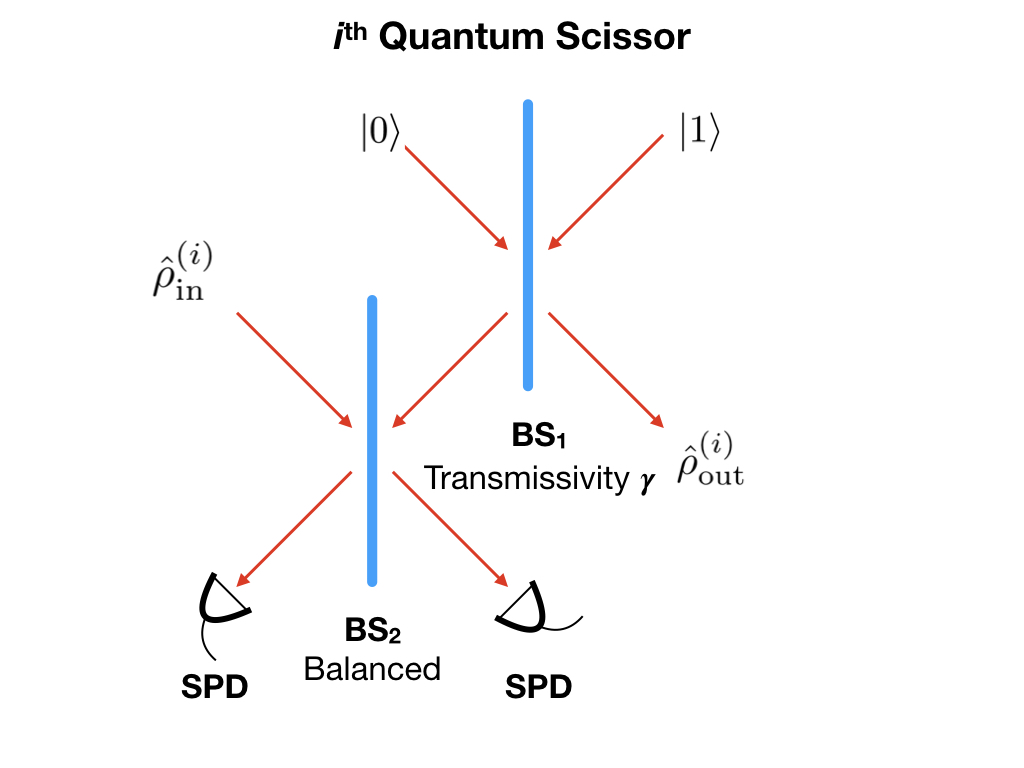}
\caption{\label{fig:NLA} The principle of noiseless linear amplification. (top) The schematic of an NLA comprised of $N$ quantum scissors (QSs) \cite{ralph2009nondeterministic}. The input quantum state $\hat{\rho}_{\rm in}$ is mixed with $N-1$ vacuum states on a $N \times N$ balanced beam splitter (BS) network. Each output mode of the BS network is individually amplified by a QS. All modes are then combined on a second $N \times N$ BS network. The output quantum state $\hat{\rho}_{\rm out}$ of the NLA is successfully generated if all quantum scissors succeed and all other output ports of the second BS network register no photons;
(bottom) Configuration of the $i^{\rm th}$ quantum scissor \cite{pegg1998optical}. BS$_1$ is an unbalanced beam splitter and BS$_2$ is a balanced beam splitter. The quantum scissor successfully amplifies the input state $\hat{\rho}^{(i)}_{\rm in}$ if either single-photon detector (SPD) registers a photon click.}
\end{figure}

The no-cloning theorem prevents a quantum state of light from being {\it deterministically} amplified without introducing noise. The no-cloning theorem, however, does not preclude {\it non-deterministic} noiseless amplifiers that only succeed with a sub-unity probability. Ralph and Lund first proposed a structured design for the noiseless linear amplifier (NLA) \cite{ralph2009nondeterministic} and showed that an input coherent states $\ket{\alpha}$ can be noiselessly amplified, with a success probability $p$, to a target coherent state $\ket{g\alpha}$, where $g$ is the amplitude gain of the NLA. Critically, a success event of the NLA is heralded so that the amplified quantum state can be utilized in subsequent quantum information processing units.  Fig.~\ref{fig:NLA} (top) illustrates a structured realization of the NLA. A BS network first diverts an input quantum state $\hat{\rho}_{\rm in}$ into $N$ spatial modes with $\ll 1$ mean photons per mode. The quantum states $\hat{\rho}^{(i)}_{\rm in}$'s are amplified by $N$ quantum scissors \cite{pegg1998optical} followed by a second BS network that coherently recombines all $N$ modes to generate the amplified quantum state $\hat{\rho}_{\rm out}$ at the NLA output. The success of the NLA is heralded only when all quantum scissors have succeeded {\it and} the single-photon detectors placed at the remaining $N-1$ NLA output modes register no photon. The $i^{\rm th}$ quantum scissor, as plotted in Fig.~\ref{fig:NLA} (bottom), consists of two BSs and two single-photon detectors. It operates on an input quantum state $\hat{\rho}^{(i)}_{\rm in}$ and an auxiliary single-photon state $\ket{1}$. BS$_1$ is unbalanced with transmissivity $\gamma$, and BS$_2$ is balanced with a $50 \%$ transmissivity. Now, let us consider a coherent-state input $\ket{\alpha}$ to an NLA with $N\gg 1$ quantum scissors. After dividing into $N$ modes, a weak coherent-state $\hat{\rho}^{(i)}_{\rm in} = \ket{\alpha_i}\bra{\alpha_i}$, i.e., $\ket{\alpha_i} = \ket{\alpha/\sqrt{N}}\simeq \ket{0} + \alpha_i \ket{1}$ and $|\alpha_i|^2 \ll 1$, is at the input to the $i^{\rm th}$ quantum scissor. A success event of the quantum scissor is heralded when {\it either} single-photon detector registers a click. Two different paths may lead to a success event: 1) the photon click originates from $\hat{\rho}^{(i)}_{\rm in}$, and the single-photon auxiliary state is reflected on BS$_1$; and 2) $\hat{\rho}^{(i)}_{\rm in}$ is its vacuum portion, and the single-photon auxiliary state transmits through BS$_1$ and results in the photon click. If the two event paths are made completely indistinguishable by perfectly matching the temporal, spectral, polarization, and spatial profiles of $\hat{\rho}^{(i)}_{\rm in}$ and the auxiliary state, their corresponding output quantum states, conditioned on a success event, need to be coherently added up to generate the output state $\hat{\rho}^{(i)}_{\rm out}$. For the input state $\hat{\rho}^{(i)}_{\rm in} = \ket{\alpha_i}\bra{\alpha_i}$, a success event yields the weak coherent state $\ket{\alpha'_i} \simeq \ket{0} + g\alpha_i \ket{1}$ at the quantum scissor output, where $g=\sqrt{(1-\gamma)/\gamma}$ The input coherent state is thus amplified by an amplitude gain of $g$. After combining the output states of all $N$ successful quantum scissors on a second BS network, the input quantum state is successfully amplified if none of the single-photon detectors at the output ports of the BS network registers a click. A success event of an ideal NLA with infinite number of quantum scissors gives
\begin{equation}
\label{eq:ideal_NLA}
    \lim_{N\to\infty} \left(1+g\frac{\alpha}{N}\hat{a}^\dagger\right)^N\ket{0} \propto \ket{g\alpha}.
\end{equation}
Hence, an input coherent state $\ket{\alpha}$ is noiselessly amplified to an output coherent state $\ket{g\alpha}$.

Because NLAs are capable of amplifying quantum states of light without introducing noise, they are potential building blocks for continuous-variable quantum repeaters \cite{dias2017quantum}. As a primitive, employing NLAs to distill the entanglement carried by a two-mode squeezed state have been both theoretically \cite{xiang2010heralded} studied and experimentally \cite{ulanov2015undoing} verified. Entanglement distillation using NLAs enables increasing the secret-key rate of QKD \cite{blandino2012improving} and the construction of continuous-variable quantum error correction codes \cite{ralph2011quantum,dias2018quantum}. More generally, Ref.~\cite{blandino2015heralded} shows that a Gaussian channel with transmissivity $\eta$ followed by an NLA with gain $g$ is equivalent to placing an effective NLA with gain 
\begin{equation}\label{eq:g_eff}
    g_{\rm eff}=\sqrt{1+(g^2-1)\eta}
\end{equation}
prior to an effective Gaussian channel with transmissivity
\begin{equation}\label{eq:effective_trans}
     \eta_{\rm eff}=\frac{g^2\eta}{1+(g^2 - 1)\eta}.
\end{equation}
Eq.~\ref{eq:effective_trans} indicates that for all $g > 1$, $\eta_{\rm eff}$ is always greater than $\eta$, suggesting that channel loss is effectively reduced by the NLA.
\begin{figure}
    \centering
    \includegraphics[width=0.5\textwidth]{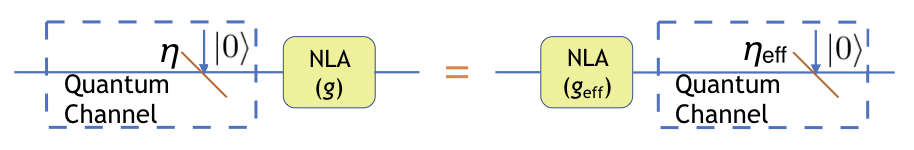}
    \caption{A NLA with gain $g$ placed after a pure loss bosonic channel with transmissivity $\eta$ is equivalent to an effective NLA with gain $g_{\rm eff}$ placed before an effective pure loss bosonic channel with tranmissivity $\eta_{\rm eff}$. $g$ and $g_{\rm eff}$ is related by Eq.~\ref{eq:g_eff}; $\eta$ and $\eta_{\rm eff}$ is related by Eq.~\ref{eq:effective_trans}.}
    \label{fig:effchannel}
\end{figure}

The ideal NLA operator $g^{\hat{n}}$, where $\hat{n}$ is the number operator, allows for noiseless phase-insensitive amplification of a coherent state, viz. $\ket{\alpha} \rightarrow \ket{g\alpha}$, but it requires infinite number of quantum scissors for its realization, as indicated by Eq.~\ref{eq:ideal_NLA}, thereby hampered by zero success probability. Apart from quantum scissors, other schemes, such as photon catalysis \cite{zhang2018photon}, photon subtraction and addition \cite{zavatta2011high}, and thermal noise addition followed by photon subtraction \cite{marek2010coherent}, have also been proposed to serve as building blocks for the NLA, but a fundamental trade-off between the success probability and the deviation from ideal noiseless amplification remains. In fact, the optimum success probability in the high-fidelity input region scales as $g^{-2N}$ \cite{pandey2013quantum}, whereas the success probability scales as $(g^2+1)^{-N}$ for quantum-scissor-based NLAs. As such, the trade-off between the success probability and amplification fidelity must be accounted for in NLA-based quantum information protocols. 

\section{\label{sec:MainResult} Distributed Quantum Sensing Enhanced by Noiseless Linear Amplifiers}
For nonlocal applications of CVMP-entanglement-based DQS, loss incurred in the distribution of entanglement to spatially separated sensor nodes quickly diminishes the measurement-sensitivity advantage over product-state-based DQS. Using the framework developed in Ref.~\cite{zhuang2018distributed}, one can easily show that the measurement-sensitivity advantage degrades from 6~dB to 0.5~dB in the presence of merely 2~dB entanglement-distribution loss. Here, we introduce NLAs into CVMP-entanglement-based DQS to tackle this problem. 

\subsection{\label{sec:DQS_idealNLA} Distributed Quantum Sensing with Ideal NLAs}
\begin{figure*}[]
\includegraphics[width=0.48\textwidth]{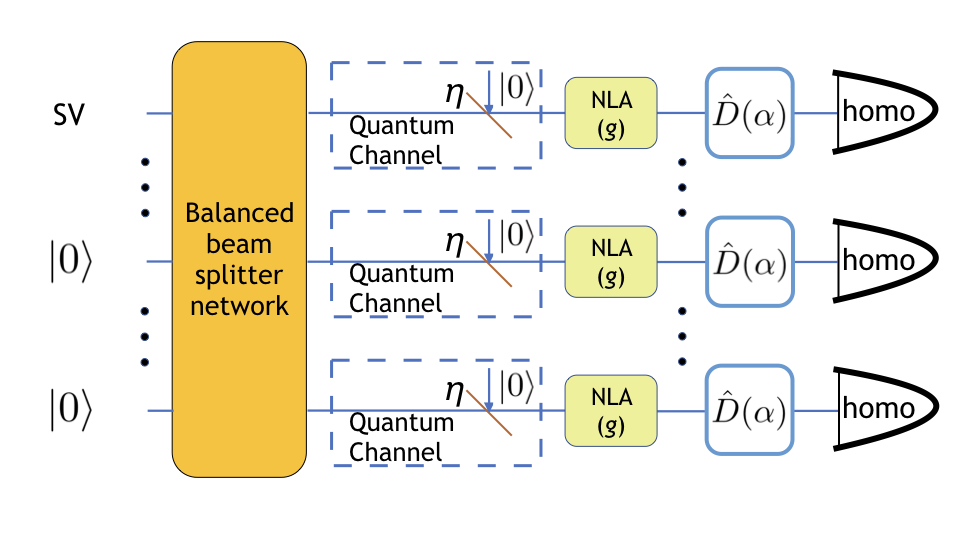}
\includegraphics[width=0.48\textwidth]{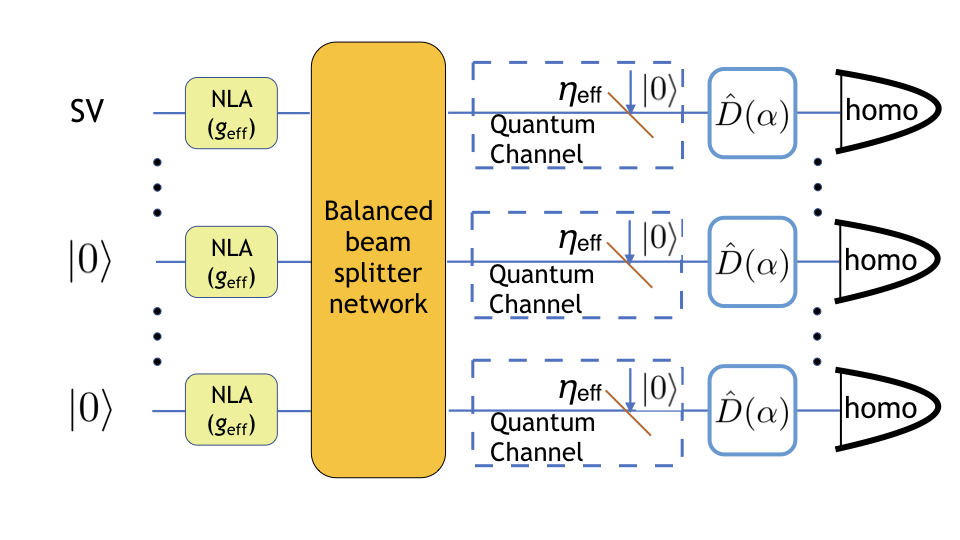}
\caption{\label{fig:DQS_NLA} Continuous-variable DQS enhanced by NLAs. (left) DQS over photon loss channel enhanced by NLAs with gain $g$. SV: squeezed vacuum state with mean photon number $N_S$; $\eta$: channel transmissivity; (right) The effective DQS scheme in which the effective NLAs are placed right after the entanglement source; $\eta_{\rm eff}$: the transmissivity of the effective channel given by Eq.~\ref{eq:g_eff}. $g_{\text{eff}}$: the gain of the effective NLA defined in Eq.~\ref{eq:effective_trans}. }
\end{figure*}

We first use ideal NLAs to illustrate the principle of NLA-enhanced DQS. Ideal NLAs offer genuine noiseless amplification, but suffer from zero success probability. We will study practical NLAs  for non-zero success probabilities in Sec.~\ref{sec:DQS_practicalNLA}. Consider the DQS scheme depicted in Fig.~\ref{fig:DQS_NLA} (left). At the source, a SV state mixes with $M-1$ vacuum states on a $M\times M$ BS network to generate the needed CVMP entanglement as the probe state. The $M$ modes of the CVMP entangled state are delivered to the sensor nodes through $M$ $\eta$-transmissivity channels. Prior to the sensing attempts, the quantum state received at each sensor node first undergoes an ideal NLA with gain $g$. Using the result reported in Ref.~\cite{blandino2015heralded}, the entanglement-distribution channel followed by the NLA is equivalent to placing an effective NLA with gain $g_{\rm eff}$ in front of an effective lossy channel with transmissivity $\eta_{\rm eff}$, where $g_{\rm eff}$ and $\eta_{\rm eff}$ are determined, respectively, by Eqs.~\ref{eq:g_eff} and \ref{eq:effective_trans}. To further move the NLAs toward the source, we now prove that the ideal NLA operator $g^{\hat{n}}$ commutes with the BS operator: starting with the commutator
\begin{equation}
    [g^{\hat{n}_a+\hat{n}_b},e^{\theta(\hat{a}^\dagger\hat{b}-\hat{a}\hat{b}^\dagger)}]= [e^{\ln(g)(\hat{n}_a+\hat{n}_b)},e^{\theta(\hat{a}^\dagger\hat{b}-\hat{a}\hat{b}^\dagger)}],
\end{equation}
it is easy to show that $[\hat{n}_a+\hat{n}_b,\hat{a}^\dagger\hat{b}-\hat{a}\hat{b}^\dagger]=0$, thus completing the proof. Using the proven commutation relation, we can now switch the order between the effective NLAs and the BS network, leading to an equivalent DQS scheme illustrated in Fig.~\ref{fig:DQS_NLA} (right). Because a vacuum state remains unchanged after being processed by an ideal NLA, we can simply eliminate all the NLAs operating on vacuum modes. We further note that an ideal NLA acting on a SV state with mean photon number $N_S$ leads to an effective SV state whose mean photon number $N_{{S_{\rm eff}}}$ is related to $N_S$ by \cite{gagatsos2012probabilistic}
\begin{equation}\label{eq:effective_SV}
    \sqrt{\frac{N_{S_{\rm eff}}}{N_{S_{\rm eff}}+1}}=g_{\rm eff}^2\sqrt{\frac{N_S}{N_S+1}}.
\end{equation}

As such, the original DQS scheme with ideal NLAs placed after the transmissivity-$\eta$ channels is equivalent to an effective DQS scheme over less lossy channels each with transmissivity $\eta_{\rm eff}$. This shows that NLAs are able to reduce the loss induced in the distribution of entanglement to spatially separated sensor nodes.

We now consider the problem of estimating the quadrature displacement, imparted by the displacement operator $\hat{D}(\alpha)$ at all sensor nodes, in DQS using CVMP entanglement and ideal NLAs. The estimator for this problem is defined as $\hat{X}_{\alpha}=\frac{1}{M}\sum_{m=1}^M \hat{X}_m$. Working with the equivalent DQS scheme and utilizing the result of Ref.~\cite{zhuang2018distributed}, the rms estimation error is readily derived to be
\begin{equation}
    \delta\alpha_{\text{NLA}}=1/2\sqrt{\frac{ \eta_{\rm eff}}{M(\sqrt{N_{S_{\rm eff}}+1}+\sqrt{N_{S_{\rm eff}}})^2}+\frac{1- \eta_{\rm eff}}{M}}.
\end{equation}
In comparison, the rms estimation error for CVMP-entanglement-based DQS without NLAs reads
\begin{equation}
    \delta\alpha_{\text{NLA}}=1/2\sqrt{\frac{\eta}{M(\sqrt{N_S+1}+\sqrt{N_S})^2}+\frac{1-\eta}{M}}.
\end{equation}
In the low brightness regime, i.e., $N_S < 1$, NLAs with gain $g \gg 1$ is physically permitted if $g_{\rm eff}^2\sqrt{\frac{N_S}{N_S+1}}<1$, yielding
\begin{equation}
    \eta_{\rm eff} \simeq \frac{g^2\eta}{1+g^2\eta} \approx 1.
\end{equation}
This result is in consistency with the loss suppression protocol introduced in Ref.~\cite{mivcuda2012noiseless}: to overcome the channel loss, the input quantum state is first noiselessly attenuated before being sent through the quantum channel and is then noiselessly amplified after transmission. In our DQS scheme, noiseless attenuation on a SV state generates a weaker SV state. As such, transmitting a CVMP entangled state with $N_S \ll 1$ through a lossy channel followed by high-gain ideal NLAs can compensate for large amount of loss. Ideal NLAs, however, have zero success probability and are thus unphysical. In the next Section we will discuss DQS with practical NLAs and discuss the trade-off between their success probability and the deviation from ideal NLAs.

\subsection{\label{sec:DQS_practicalNLA} Distributed Quantum Sensing Enhanced by Practical NLAs}
A practical NLA, one with finite success probability at the cost of introducing a small amount of amplification noise, is comprised of $N$ quantum scissors that truncate the Hilbert space. The practical NLA operator is written as $\hat{T}=\hat{\Pi}_N g^{\hat{n}}$, where the projection operator $\hat{\Pi}_N$ is defined as
\begin{equation}\label{eq:projection}
    \hat{\Pi}_N=\left(\frac{1}{g^2+1}\right)^{\frac{N}{2}}\sum_{n=0}^N \frac{N!}{(N-n)!N^n}\ket{n}\bra{n}.
\end{equation}

As the number of quantum scissors increases, a practical NLA approaches an ideal one. Applying a practical NLA on an input quantum state yields the output quantum state \cite{dias2017quantum}
\begin{equation}
    \hat{\rho}_{\rm out}=\frac{\hat{T}\hat{\rho}_{\rm in}\hat{T}}{{\rm Tr}(\hat{T}\hat{\rho}_{\rm in}\hat{T})}
\end{equation}
with a success probability
\begin{equation}
    p={\rm Tr}(\hat{T}\hat{\rho}_{\rm in}\hat{T}).
\end{equation}
The measurement sensitivity of using the amplified state $\hat{\rho}_{\rm out}$ as the probe is thus given by
\begin{equation}
    {\rm Var}(\hat{X}_{\alpha})= {\rm Tr}(\hat{\rho}_{\rm out} \hat{X}_{\alpha}^2)-{\rm Tr}(\hat{\rho}_{\rm out} \hat{X}_{\alpha})^2.
\end{equation}

\begin{figure}[tbh!]
 \includegraphics[width=0.49\textwidth]{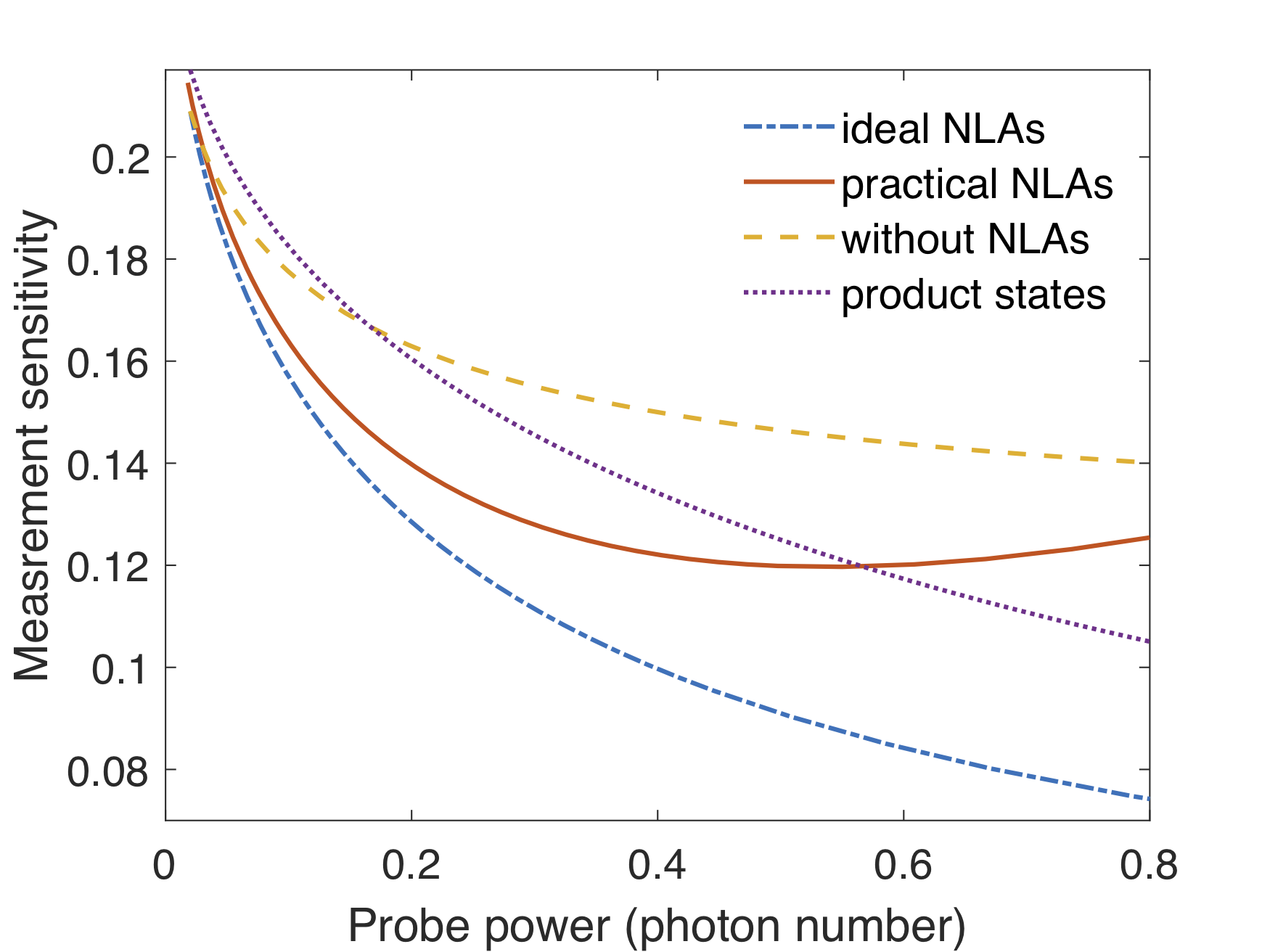}
 \caption{\label{fig:pNLA_sensitivity} Measurement sensitivity versus the probe power for DQS using CVMP entanglement and ideal NLAs (dash-dot curve); CVMP entanglement and practical NLAs with two quantum scissors (solid curve); CVMP entanglement without using NLAs (dashed curve); and the optimum product states (dotted curve). The transmissivity for entanglement-distribution channels is set to $\eta=0.5$.}
\end{figure}

\begin{figure}
  \includegraphics[width=0.49\textwidth]{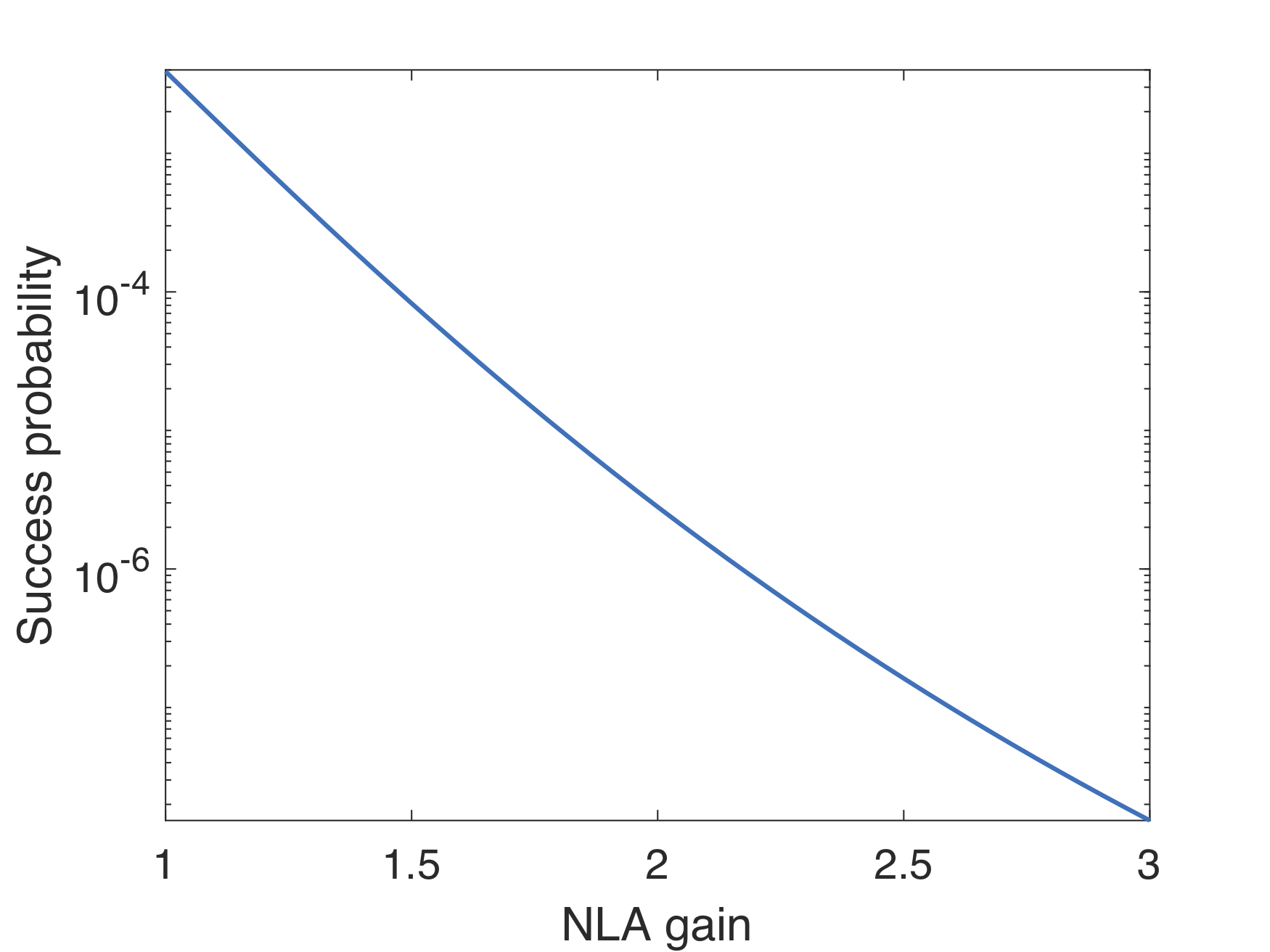}
  \includegraphics[width=0.49\textwidth]{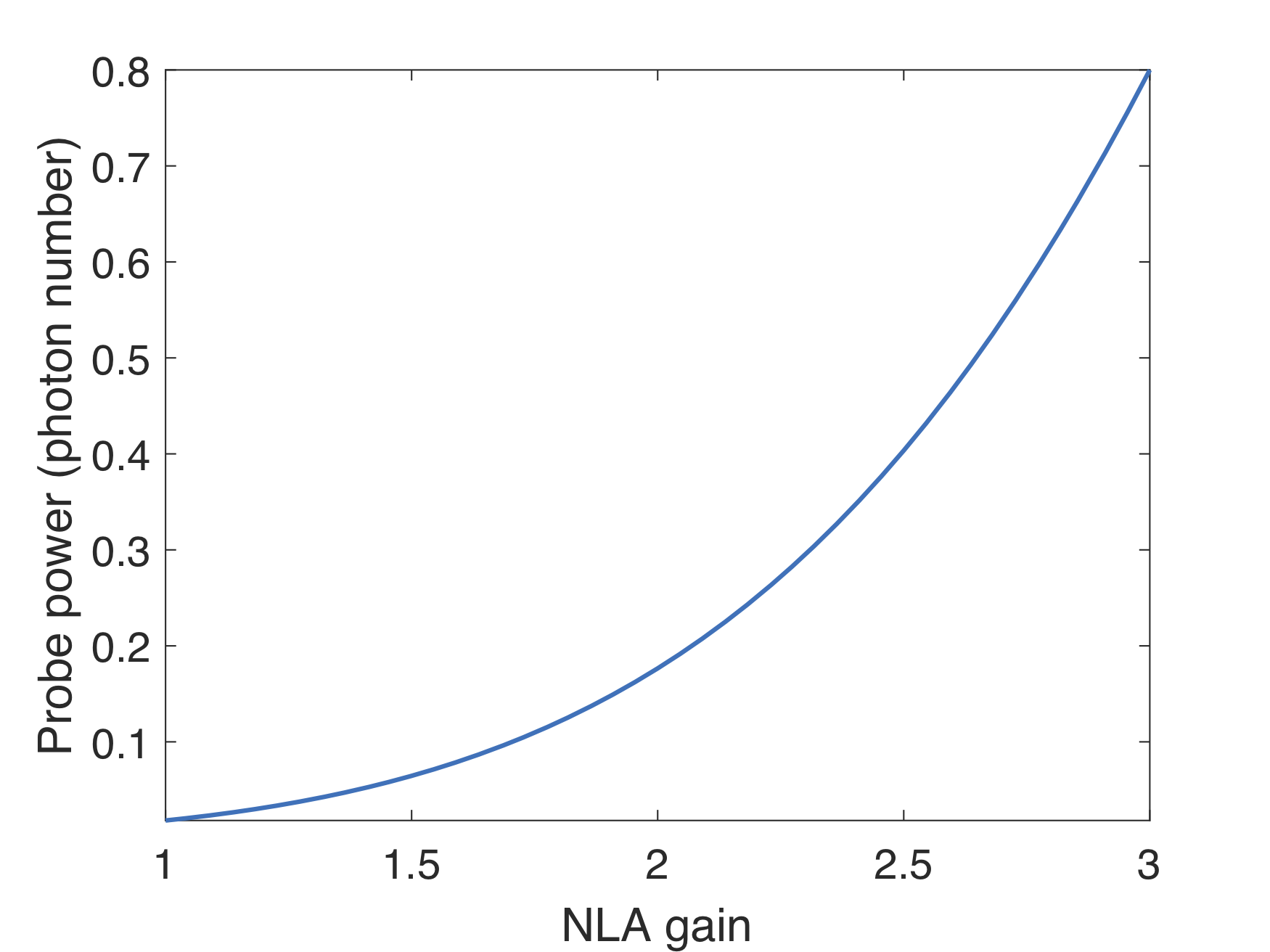}
\caption{\label{fig:pNLA_success} (top) The success probability of practical NLA at different gain. (bottom) The probe power after performing practical NLAs at different gain.}
\end{figure}

The measurement sensitivity at different probe power levels for four DQS scenarios, including CVMP entanglement and ideal NLAs, CVMP entanglement and practical NLAs, CVMP entanglement without NLAs, and the optimum product state, is plotted in Fig.~\ref{fig:pNLA_sensitivity}. The probe power for CVMP entanglement with ideal NLAs is given by $N_P = N_{S_{\rm eff}} \eta_{\rm eff}$ and for CVMP entanglement without NLAs is given by $N_P = N_S \eta$.  Here, we consider a four-node DQS situation with $N_S = 0.04$, and channel transmissivity $\eta = 0.5$. As shown in the figure, ideal NLAs, despite having a zero success probability, allow for the best measurement sensitivity (dash-dot curve) among all four scenarios. To obtain a finite success probability, we next analyze the measurement-sensitivity achieved by using practical NLAs each with two quantum scissors, in lieu of an ideal NLA, at each sensor node. Due to the excess noise of practical NLAs, their achieved measurement sensitivity (solid curve) is worse than that enabled by ideal NLAs, but still outperforms that of CVMP-entanglement-based DQS without NLAs (dashed curve). The three measurement-sensitivity curves are compared to the one achieved by using the optimum product states, i.e., $M$ SV states, with the same probe power (see Appendix \ref{appendix:opt_proof} for a proof for the optimality of product SV states). At a probe power level between $\sim$ 0.18 and $\sim$ 0.58 photons, the measurement sensitivity enhanced by CVMP entanglement in conjunction with practical NLAs beats the measurement sensitivity achieved by the optimum product states, but no such performance advantage is attainable without NLAs. Thus, the NLAs provide the essential functionality desired by quantum repeaters –– regeneration of entanglement to enable performance superior to what the optimum local protocol affords. As $N_P$ increases, more NLA gain is needed, resulting in a degradation in the measurement sensitivity at $N_P > 0.5$. This suggests that more quantum scissors are needed to reduce the excess noise introduced during the high-NLA-gain amplification process. However, the success probability will decrease exponentially as it scales as $\propto (\frac{1}{g^2+1})^N$, where $N$ is the number of quantum scissors. 

As discussed at the outset, a fundamental trade-off between the success probability of practical NLAs and the amount of excess amplification noise exists. Fig.~\ref{fig:pNLA_success} (top) plots the success probability for the practical NLAs at different gains, whose corresponding probe power is depicted in Fig.~\ref{fig:pNLA_success} (bottom). At a probe power level $\sim$ 0.2 photons, the measurement sensitivity advantage enjoyed by CVMP entanglement with practical NLAs over that achieved by the optimum product states is maximized. The corresponding gain of each NLA is 2.2 and a probability of $10^{-5}$ for {\it all} NLAs to succeed. The bandwidth for narrowband CVMP entangled states produced in optical cavities is typically in the rage of a few hundred mega Hertz to a few giga Hertz. Hence a $10^{-5}$ success probability leads to a few thousand success events per second.  



\section{\label{sec:Discussions}Discussions}
In general, practical NLAs provide an appreciable measurement-sensitivity improvement in the low photon-number regime, i.e., $N_S \ll 1$, where the effective transmissivity becomes $\eta_{\rm eff} \sim 0.86$ at NLA gain $g\sim2.5$, source brightness $N_S = 0.04$, and $\eta = 0.5$. 

In the large entanglement-distribution loss regime, e.g., $\eta<0.4$, the optimum product states afford a comparable measurement sensitivity than that achieved by CVMP entanglement and practical NLAs, because product states can be generated locally without being plagued by entanglement-distribution loss. In the low entanglement-distribution loss regime, e.g., $\eta>0.7$, the quantum sensors receive more photons than they do in the large entanglement-distribution loss regime. The success probability for NLAs will consequently drop from that in the high entanglement-distribution loss regime when choosing the same gain, resulting in a limited measurement sensitivity improvement. Also, the effective channel transmissivity $\eta_{\rm eff} \approx \frac{g^2\eta}{1+g^2-1}=\eta$ for $\eta \sim 1$. In this case, practical NLAs do not offer a significant measurement-sensitivity advantage over that obtained by CVMP entanglement without NLAs.

The success probability of practical NLAs applied on coherent states is upper bounded by  $1/g^{2N}$ in the high fidelity input region \cite{pandey2013quantum}. The associated NLA operator $g^{-N} g^{\hat{n}}$ $\hat{P}_N$, where $\hat{P}_{N}$ determines the cutoff in the number basis. The success probability of the practical NLA based on the projector operator defined in Eq.~\ref{eq:projection} is far below Ref.~\cite{pandey2013quantum}'s upper bound. To boost the success probability, Ref.~\cite{yang2013improving} shows that injecting SV states into the quantum scissors offers marginal improvement. An open problem is to explore alternative NLA structures to approach the success probability upper bound.

\section{\label{sec:conclusion}Conclusion}
In summary, we have studied the use of NLAs to improve the measurement sensitivity of CVMP-entanglement-based DQS over lossy entanglement-distribution channels. The trade-off between the measurement-sensitivity improvement and the success probability was studied. Our result shows that NLAs in such a sensing scenario behave like quantum repeaters that mitigate entanglement-distribution loss so that entanglement enables a measurement-sensitivity advantage over the DQS scheme based on the optimum separable product states. The practical NLAs discussed in this paper are realizable by available technology. As such, the proposed scheme may lead to the first instance of repeater-enhanced quantum sensing.\\

\begin{acknowledgments}
Z.Z. thank support from the University of Arizona. Z.Z. and W.C. acknowledge Sponsored Research Agreement No.~18-BOA-SC-0003 by General Dynamics Mission Systems. Y.X. is supported by the Nicolaas Bloembergen Graduate Student Scholarship. Q.Z. acknowledges the U.S. Department of Energy through Grant No.~PH-COMPHEP-KA24.
\end{acknowledgments}
\appendix

\section{Proof for the Optimum States for Distributed Quantum Sensing}
\label{appendix:opt_proof}

In this section, we show that in the absence of loss, (1) the CVMP entangled state generated by equally splitting a squeezed vacuum state is optimum and (2) the product of single-mode squeezed state is the optimum product state for DQS of displacements.

The displacement unitary operator $\hat{U}\left(\alpha\right)$ with $\alpha>0$ can be written as
$\hat{U}\left(\alpha\right)=\exp\left(-i \alpha \hat{p}\right)$, and $\partial_\alpha\hat{U}\left(\alpha\right)=-i \hat{p} \hat{U}\left(\alpha\right)$.
The quantum Cram\'{e}r-Rao bound~\cite{Helstrom_1976,Holevo_1982,Yuen_1973} gives the minimum rms error for estimating a single parameter $\alpha$ of a quantum state $\hat{\rho}\left(\alpha\right)$:
\begin{equation}
\delta \alpha_\eta \ge  \delta \alpha^{\rm CR} \equiv \sqrt{1/I_F[\hat{\rho}\left(\alpha\right)]},
\label{CR_bound}
\end{equation}
where the quantum Fisher information is given by
\begin{eqnarray}
I_F[\hat{\rho}(\alpha)]\equiv  \lim_{\epsilon\to 0} 8\!\left\{1-
\sqrt{{\mathcal F}[\hat{\rho}(\alpha),\hat{\rho}(\alpha+\epsilon)]}\right\}/{\epsilon^2},
\label{Uhlmann}
\end{eqnarray}
where ${\mathcal F}(\hat{\sigma}_1,\hat{\sigma}_2)\equiv \left[{\rm Tr}\!\left(\sqrt{\sqrt{\hat{\sigma}_1}\hat{\sigma}_2\sqrt{\hat{\sigma}_1}}\right)\right]^2$ is the fidelity between $\hat{\sigma}_1$ and $\hat{\sigma}_2$.  

Suppose loss changes the joint state across all nodes to $\hat{\rho}_{M,N_S,\eta}=\sum_\ell c_\ell \ket{\psi_\ell}\bra{\psi_\ell}$, where $\sum_\ell c_\ell=1$.

Because the sensing operation $\hat{U}\left(\alpha\right)$ is unitary, by the convexity of quantum Fisher information we have
\begin{equation}
I_F[\hat{\rho}(\alpha)]\le \sum_\ell c_\ell I_F[\hat{U}\left(\alpha\right)^{\otimes M}\ket{\psi_\ell}].
\label{F_convex}
\end{equation}
For a pure state, we can easily show that the fidelity
${\mathcal F}(\hat{U}\left(\alpha\right)^{\otimes M}\ket{\psi_\ell},\hat{U}\left(\alpha+\epsilon\right)^{\otimes M}\ket{\psi_\ell})
= |\braket{\psi_\ell|\hat{U}\left(\epsilon\right)^{\otimes M}|\psi_\ell}|^2
$, and thus taking derivatives gives
\begin{equation}
I_F[\hat{U}\left(\alpha\right)^{\otimes M}\ket{\psi_\ell}]
=4 {\rm var} \left(\sum_k {\hat{p}_k^\prime} \right)_{\ket{\psi_\ell}},
\end{equation}
where ${\hat{p}_k^\prime}$ is the momentum quadrature of the mode on node $k$ after loss. Combining Eq.~\ref{F_convex}, we have
\begin{eqnarray}
I_F[\hat{\rho}(\alpha)]/4 &\le& \left<\left(\sum_k\hat{p}_k^\prime \right)^2\right>_{\hat{\rho}(\alpha)}-\sum_\ell c_\ell\left<\left(\sum_k \hat{p}_k^\prime\right)\right>^2_{\ket{\psi_\ell}}\notag
\\
&\le& {\rm var}\left(\sum_k{\hat{p}_k^{\prime}}\right)_{\hat{\rho}(0)},
\end{eqnarray}

where we have used the inequality $\sum_\ell c_\ell x_\ell^2\ge \left(\sum_\ell c_\ell x_\ell\right)^2$ and the invariance of quadrature variance under displacement. 
Because the pure loss channel transforms $\hat{p}_k^\prime=\sqrt{\eta}\hat{p}_k+\sqrt{1-\eta} \hat{p}_{e_k}$, where $\hat{p}_{e_k}, \hat{p}_k$ are the momentum quadrature of the environment and the mode before loss, we immediately have
\begin{equation}
I_F[\hat{\rho}(\alpha)]/4 \le  \eta {\rm var}\left(\sum_k {\hat{p}_k} \right)+\left(1-\eta\right)M,
\end{equation}
where the variance is evaluated before loss.
Combining with ineq.~\ref{CR_bound} we have
\begin{align}
&\delta \alpha^{\rm CR}=\frac{1}{2\sqrt{\eta {\rm var}\left(\sum_k {\hat{p}_k} \right)+\left(1-\eta\right)M}}\notag
\\
&\ge\frac{1}{2}\left(\frac{1}{M\eta\left(\sqrt{N_S+1}+\sqrt{N_S}\right)^2+M\left(1-\eta\right)}\right)^{1/2}.
\end{align}
We find that at $\eta=1$, the new bound matches $\delta \alpha_\eta^E$ and thus the CVMP entangled state is optimum in absence of loss. However, when $\eta<1$, the new bound deviates from $\delta \alpha_\eta^E$, so we expect it to be not tight.

Likewise, for the product-state case, we have derived the bound
\begin{align}
&\delta \alpha^{\rm CR}\ge 
\nonumber
\\ &\frac{1}{2}\left(\frac{1}{M\eta\left(\sqrt{N_S/M+1}+\sqrt{N_S/M}\right)^2+M\left(1-\eta\right)}\right)^{1/2}.
\end{align}
We compare the above new bound with the performance of the product state scheme $\delta \alpha_\eta^P$ and find them agree at $\eta=1$, thereby concluding that a product state comprised of identical squeezed vacuum states is the optimum in the absence of loss. When $\eta<1$, the new bound deviates from $\delta \alpha_\eta^P$, so it is expected be be loose.

Note that the approach described in Ref.~\cite{Escher_2011} gives the same bounds for the minimum rms estimation error.

\end{document}